\documentclass{article}

\usepackage{microtype}
\usepackage{graphicx}
\usepackage{subfigure}
\usepackage{hyperref}
\usepackage{gensymb}

\usepackage{amsmath}
\DeclareMathOperator{\atantwo}{atan2}

\DeclareMathOperator{\trace}{tr}

\usepackage[accepted]{icml2021}

\icmltitlerunning{Machine Learning Approach to Determine Orientation Fields}

\begin{document}

\twocolumn[
\icmltitle{A Machine Learning Approach to Robustly Determine Director Fields and Analyze Defects in Active Nematics}



\icmlsetsymbol{equal}{*}

\begin{icmlauthorlist}
\icmlauthor{Yunrui Li}{cs}
\icmlauthor{Zahra Zarei}{phy}
\icmlauthor{Phu N. Tran}{cs, phy}
\icmlauthor{Yifei Wang}{cs}
\icmlauthor{Aparna Baskaran}{phy}
\icmlauthor{Seth Fraden}{phy}
\icmlauthor{Michael F. Hagan}{phy}
\icmlauthor{Pengyu Hong}{cs}
\end{icmlauthorlist}

\icmlaffiliation{cs}{Computer Science Department, Brandeis University, USA}
\icmlaffiliation{phy}{Physics Department, Brandeis University, USA}

\icmlcorrespondingauthor{Pengyu Hong}{hongpeng@brandeis.edu}
\icmlcorrespondingauthor{Yunrui Li}{yunruili@brandeis.edu}

\icmlkeywords{Machine Learning, ICML}

\vskip 0.3in
]




\begin{abstract}
Active nematics are dense systems of rodlike particles that consume energy to drive motion at the level of the individual particles. They exist in natural systems like biological tissues and artificial materials such as suspensions of self-propelled colloidal particles or synthetic microswimmers. Active nematics have attracted significant attention in recent years due to their spectacular nonequilibrium collective spatiotemporal dynamics, which may enable applications in fields such as robotics, drug delivery, and materials science. The director field, which measures the direction and degree of alignment of the local nematic orientation, is a crucial characteristic of active nematic and is essential for studying topological defects. However, determining the director field is a significant challenge in many experimental systems. Although director fields can be derived from images of active nematics using traditional imaging processing methods, the accuracy of such methods are highly sensitive to the settings of the algorithms. These settings must be tuned from image-to-image due to experimental noise, intrinsic noise of the imaging technology, and perturbations caused by changes in experimental conditions. This sensitivity currently limits automatic analysis of active nematics. To address this, we developed a machine learning model for extracting reliable director fields from raw experimental images, which enables accurate analysis of topological defects. Application of the algorithm to experimental data demonstrates that the approach is robust and highly generalizable to experimental settings that are different from those in the training data. It could be a promising tool for investigating active nematics and may be generalized to other active matter systems.

\end{abstract}

\newcommand*{\Loss}{\text{Loss}}

\section{Introduction}

\begin{figure*}
    \centering
    \includegraphics[width=0.8\textwidth]{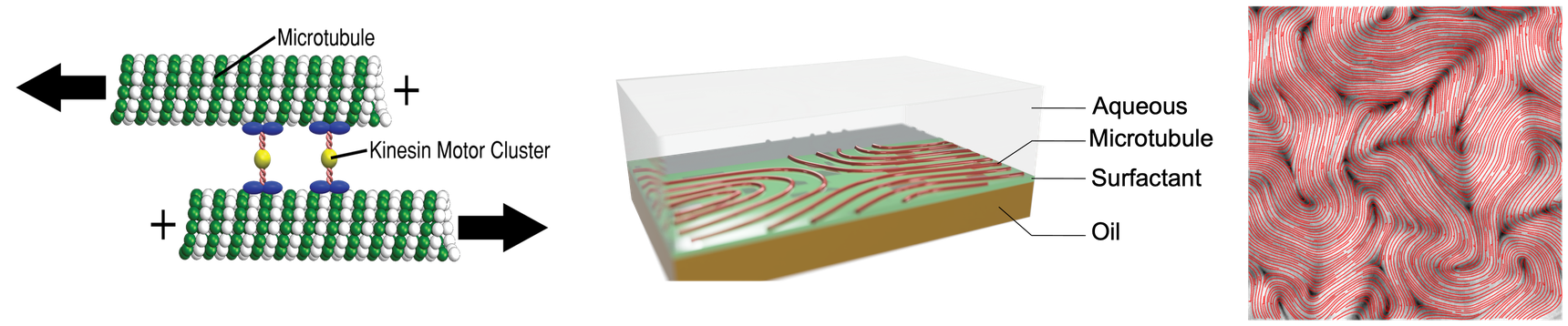}
    \caption{\textbf{Left}: The experimental data in this work is obtained from active nematics that consist of microtubule bundles and kinesin molecular motors that bind multiple microtubules. The kinesins are powered by ATP to walk along microtubules. When kinesins bound to antiparallel microtubules walk toward their respective positive ends, they can cause the microtubules to slide relative to each other, contributing to the dynamics observed in active nematic systems.  \textbf{Middle}: Experimental set-up of 2D active nematics, in which the materials are placed at an oil-water interface. \textbf{Right}: A director field (red curves, captured by PolScope) that is superimposed on the corresponding retardance image.} 
    \label{activefilament}
\end{figure*}

Active materials are ubiquitous non-equilibrium systems that can be found in many natural processes and biological systems, such as cellular cytoskeletons \cite{saw2017topological, kawaguchi2017topological, prost2015active}, bacterial suspensions \cite{sokolov2007concentration, wensink2012meso, dunkel2013fluid}, flocks of birds \cite{toner1995long},  and schools of fish \cite{katz2011inferring}. The components of such systems consume energy to drive forces or motion at the scale of individual particles, which leads to emergent macroscale collective dynamics \cite{marchetti_hydrodynamics_2013, ramaswamy2010mechanics}. In this work, we study a class of active matter known as an active nematic, which describes a system of rodlike particles that are placed at sufficient density to form a nematic liquid crystal with orientational order, but are continually driven out of equilibrium by propulsion of the particles. We focus on a widely studied experimental model active nematic system, consisting of elongated bundles of microtubules and molecular motors (kinesins), in which the kinesins consume adenosine triphosphate (ATP) fuel to slide microtubules relative to each other  (Figure \ref{activefilament}) \cite{decamp2015orientational, sanchez2012spontaneous, henkin2014tunable}. These motions produce topological defects and chaotic flows that compete with the equilibrium orientational order, producing spectacular spatiotemporal behaviors. Describing and understanding the rich behaviors that arise from this interplay is a challenge at the forefront of modern physics.

A remarkable characteristic of active nematics is the continuous proliferation and annihilation of topological defects in regions where the nematic orientational order is disrupted \cite{putzig2016instabilities, tan2019topological}. This can occur when the activity of the particles in the nematic is sufficiently high to give rise to a flow that deforms the nematic structure, resulting in the formation of defects. Topological defects take a variety of forms and are important for a number of reasons. Firstly, they can serve as a measure of the activity of the nematic, with a higher density of defects indicating a higher level of activity. Secondly, defects can act as sources or sinks for the flow of the nematic, influencing the overall flow patterns and dynamics of the system. Finally, defects can also affect the transport properties of the nematic, such as its viscosity or diffusivity, which can have important implications for the behavior of the nematic as a whole. Overall, the study of topological defects in active nematics is a rich and active area of research, with many interesting questions still remaining to be answered \cite{serra2023defect, ignes2023active, prost2015active, hagan2016emergent, bar2020self}. One important way to understand the defects is from the director fields of a given system. The director field in active nematics is crucial in detecting defects because it represents the local direction of motion of the active particles and provides a way to quantify the degree of orientational order in the material. Defects in active nematics correspond to discontinuities in the orientational order; gradients in the magnitude and direction of the surrounding director field can be used to identify defects and track their motions.

\subsection{Existing methods to obtain director fields}

\begin{figure}
    \centering
    \includegraphics[width=0.95\linewidth]{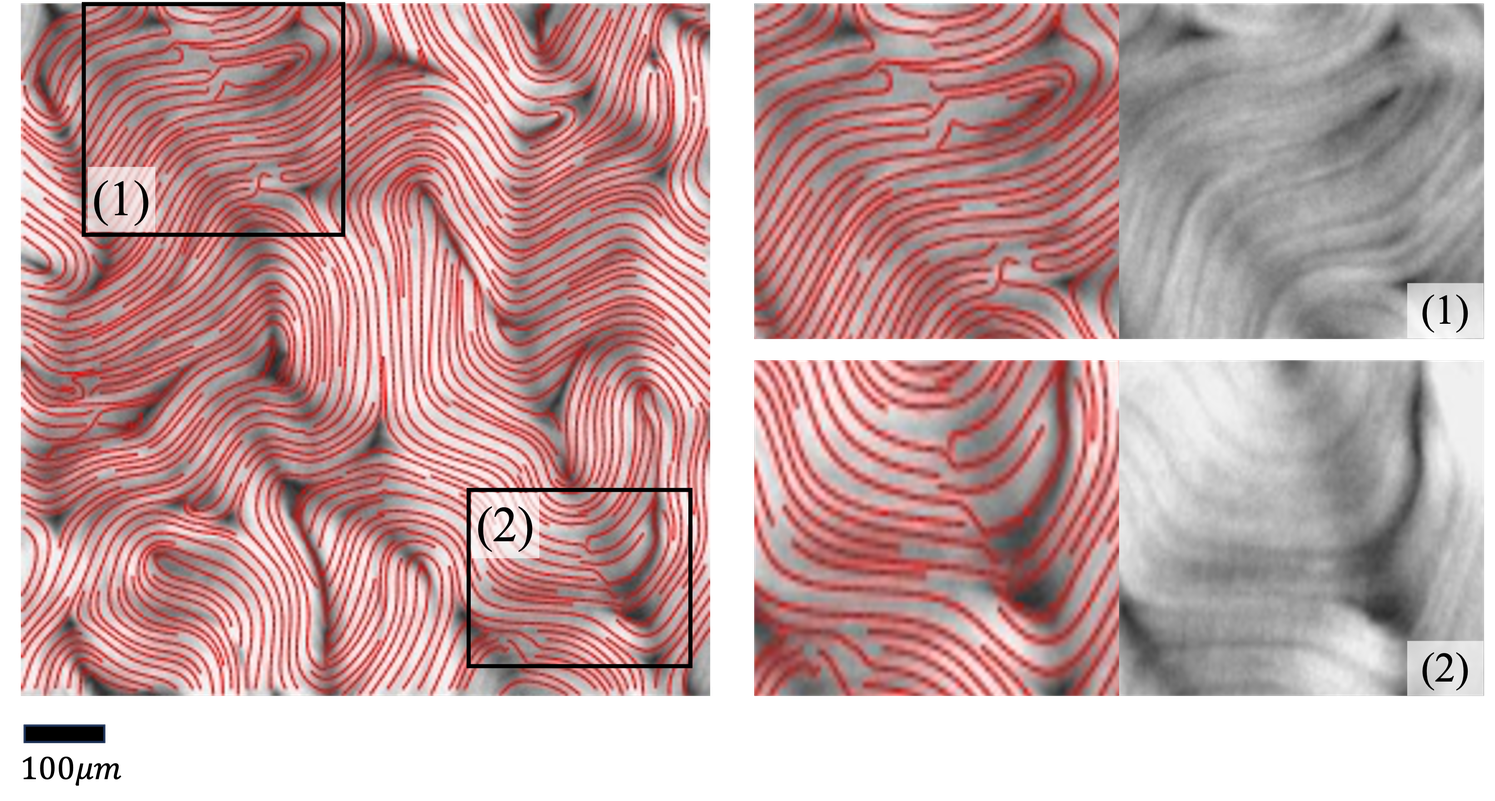}
    \caption{Examples of discontinuous director fields obtained by PolScope. In the highlighted areas (1) and (2), the calculated director fields contain discontinuities, while the retardance images show aligned microtubule bundles.}
    \label{fig:unsmooth}
\end{figure}

The nematic tensor, or the $Q$-tensor \cite{bantysh2023first, golden2023physically, doostmohammadiactive2018, keber2014topology, ramaswamy2010mechanics, doostmohammadi2016stabilization}, describes the director field; that is, the degree of alignment and the direction of the local nematic orientation. A reliable director field is essential to accurately detect, measure, and track defects in active nematic systems. One way to calculate the director field from experiments is to use the  PolScope \cite{oldenbourg2005polarization, zhou2014living, shribak2003techniques},  which is a powerful tool for visualizing and analyzing the dynamics of nematic structures. It operates by shining polarized light through the specimen, which interacts with the aligned molecules or particles in the nematic phase. This interaction alters the polarization of the light, allowing the PolScope to measure birefringence quantitatively at every single pixel in the image. The resulting retardation of light allows the PolScope to accurately determine the degree and orientation of microtubule alignment, pixel by pixel, providing important insights into the structural characteristics and behavior of liquid crystal systems. PolScope generates a pair of images - the retardance image and the corresponding director field image, at any given time point.  The calculated director field enables a direct comparison between experimental observations and theoretical results. This data also can be used to train dynamics prediction models, such as proposed by \citet{zhou2021machine,colen2021machine}.
Nevertheless, certain experimental setups and controls are not feasible using PolScope. For example, some experiments use light to control the dynamics of active nematics \cite{zarei2023light,zhang2021spatiotemporal, lemma2022spatiotemporal, ross2019controlling}. These experiments illuminate the sample from above, which
makes the traditional PolScope setup impractical. In addition, high numerical aperture condenser and objective lenses typically introduce distortions that limit the purity of polarization in a light microscope. These distortions present challenges and limitations when using a
PolScope \cite{oldenbourg2005polarization}. Lastly, the nematic material's small retardance values are often overshadowed by stronger background birefringence from microscope objective strain and Fresnel reflectivity. Although background correction is a standard practice, it doesn't fully account for the microscope slide's birefringence and is sensitive to variations across the image, optical alignment, and temperature. This can result in inaccuracies in local orientation measurements in some regions. For example,  discontinuous director fields were observed in the middle of microtubule bundles, see Figure \ref{fig:unsmooth}. This may result in false detection and inaccurate localization of defects. 

Alternatively, other instruments can offer more flexible control and design of the experiments, but cannot generate the corresponding director fields directly. For example, a fluorescence microscope enables direct light control \cite{zhang2021spatiotemporal, lemma2022spatiotemporal, ross2019controlling} or ATP control of the system's activities, but does not offer the capability to extract director fields.  In some experiments, specialized fluorescent dyes or proteins are added at low concentrations to enable the visualization and tracking of individual microtubule movements. However, such methods only provide sparse director field information that is randomly distributed across the whole view.
In these situations, one can use image processing methods to extract orientations from raw images directly from experiments, such as fluorescence images. Usually, an in-house method is developed using softwares such as MATLAB for extracting director fields from raw images \cite{tan2019topological, shi2013topological, blackwell2016microscopic, duclos2020topological, rezakhaniha2012experimental}. The methods follow a similar approach by first finding the gradient change of a raw image, then coarse-graining the result, and finding the smallest eigenvalue. More details are presented in Appendix A.1 and \citet{ellis2018curvature}. In this work, we use one of the existing implementations of traditional methods developed by Michael M. Norton\protect\footnote[3]{Relevant code can be found on GitHub: \url{https://github.com/wearefor/qcon_nematicdefectfinder}},  referred to as the ``TM'' in the rest of this paper.

\begin{figure}
    \centering
    \includegraphics[width=0.95\linewidth]{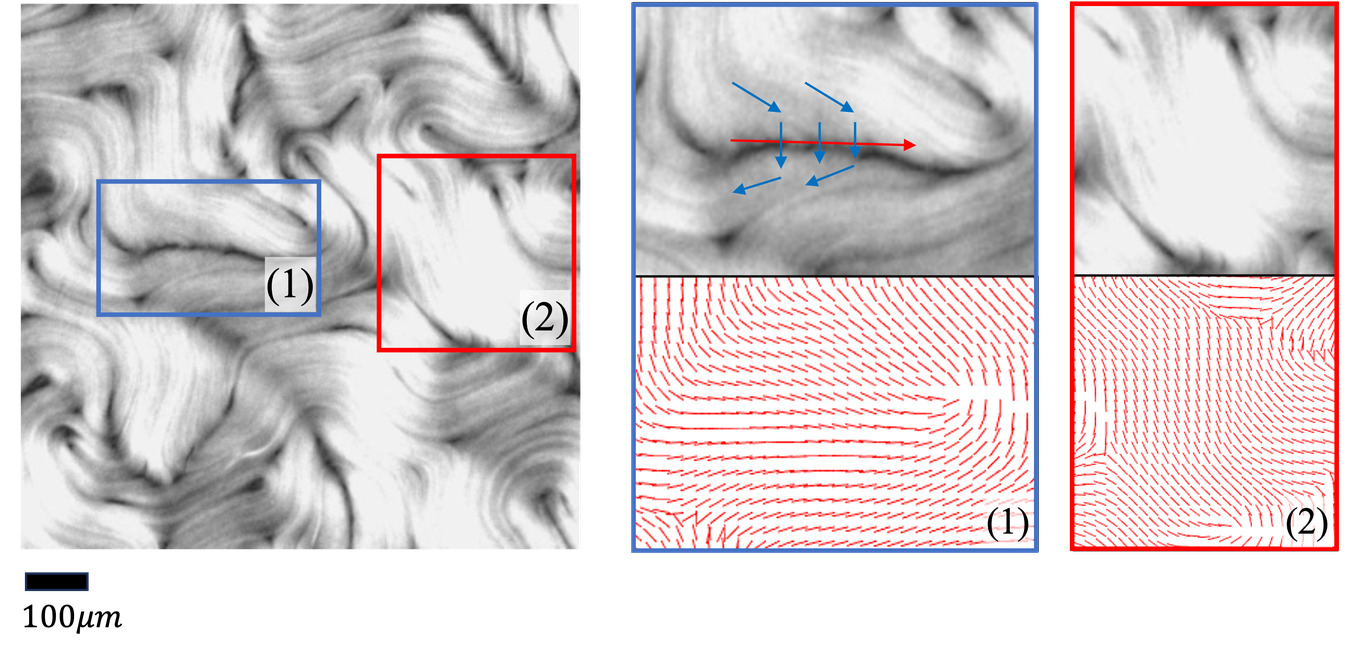}
    \caption{Challenges of extracting director field information from retardance images. \textbf{Left:} An exemplar retardance image with complex regions highlighted. \textbf{Right:} Zoomed-in regions (top) and their director fields calculated by the TM (bottom). (1)  Microtubule bundles are bent in sharp U-shapes to form an elongated positive defect, creating a region with a low density of microtubule bundles that exposes the container base and is hence visible as a dark horizontal ``crack'' across the U-shaped bundles. This ``crack'' highlights a sharp contrast with the surrounding microtubule arrangement. The blue arrows indicate the rough trends of the correct orientations calculated by PolScope (which we use as the ground truth), while the red arrow indicates the direction of the ``crack''. Due to the sharp contrast between the ``crack'' and its neighborhood, the TM may erroneously identify the crack as a collection of horizontally oriented microtubule bundles and estimate the local orientation of the region to be horizontal. However, in fact, the local orientation of the microtubule bundles at the ``crack'' region should be vertical. (2) An over-exposed region due to fluctuations of the material in the vertical direction (along the z-axis). The TM estimates the local orientation to be nearly uniform throughout the area.}
    \label{noise_ret}
\end{figure}

 However, the effectiveness of the TM can be limited because raw images contain complex signals arising from variations in experimental settings, noise, limitations of microscopy, etc. Examples of such limitations include the following. First, the analysis of retardance images is intrinsically difficult at regions that have low contrast, are out-of-focus, or have distortions due to a variety of effects. For example, although active nematics are assumed to be two-dimensional by imaging systems, they actually move in a shallow three-dimensional container.  This three-dimensionality can influence both the illumination and the focus of the microscope, thereby altering the pixel intensity in the resultant images. Additionally, active nematic filaments, when compressed by adjacent ones, can buckle into a U shape, causing a fissure that exposes the base of the container. Such fissures often run counter to the direction of movement, which can result in a miscalculation of the local orientation. We show some examples of orientation miscalculation by the TM in noisy regions in Figure \ref{noise_ret}. Finally,  the inherent variability in active nematics systems and the varying levels of activity observed across different experiments contribute to a diverse distribution of noisy regions across successive frames. 

\begin{figure*}[!h]
    \centering
    \includegraphics[width=0.95\textwidth]{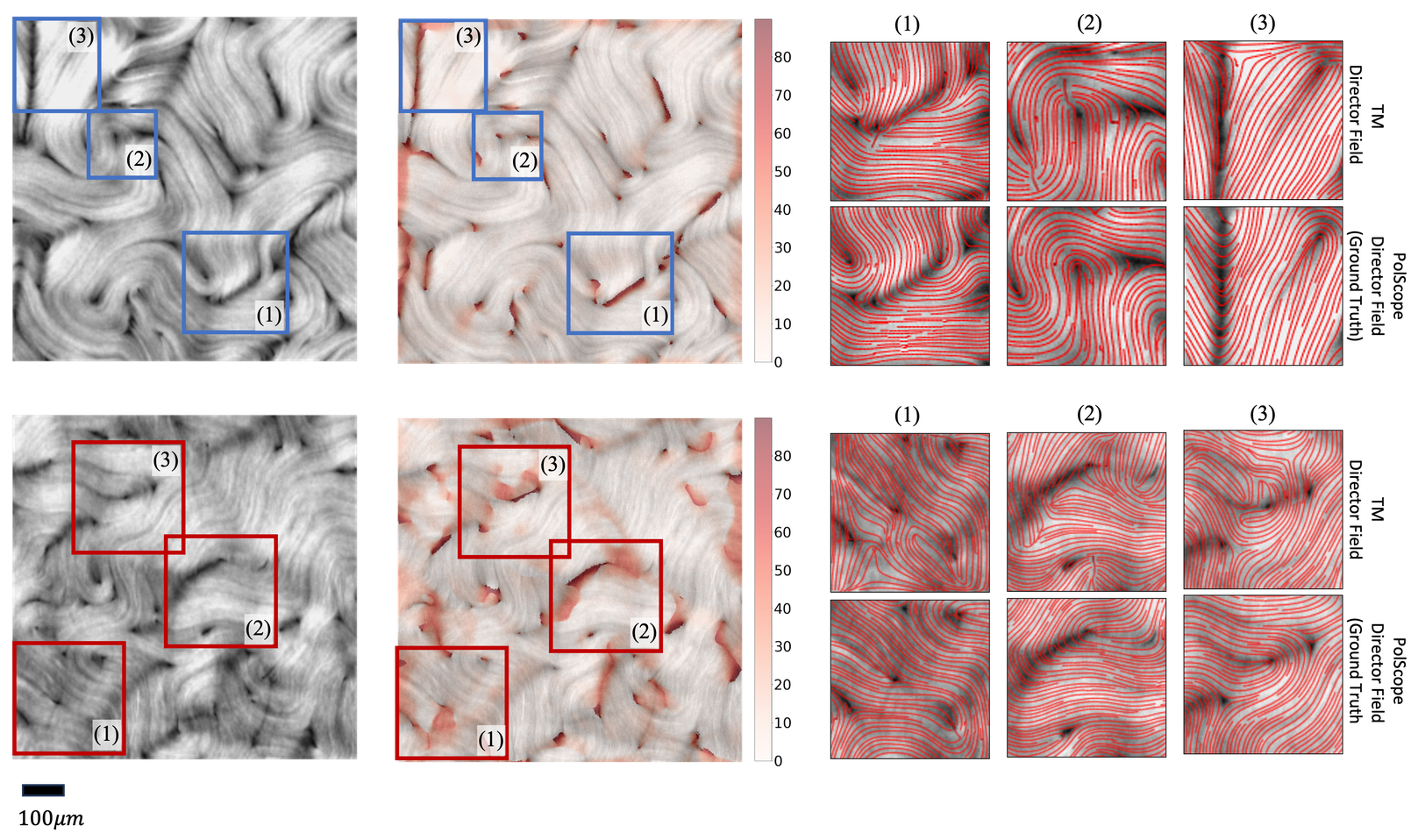}
    \caption{Examples of the director fields extracted from retardance images using the (TM) method. The top and bottom rows show the calculated director fields of two retardance images from the same video at different times. The parameter settings of the TM were manually optimized for the top image and were then applied to the bottom image. \textbf{Left column}: Retardance images. \textbf{Middle column}: The calculated director fields are compared with those obtained by PolScope  (which we use as ground truth). The differences are marked in red (the darker the bigger the difference,  measured in degrees ($\degree$)). \textbf{Right column}: Zoom-in views of the areas (1, 2, 3) marked in the First and Middle columns. The calculated local orientation is fairly accurate in the earlier frame but is much less accurate in the later frame.  $\sigma$ in the TM is set to 10.}
    \label{comporientation}
\end{figure*}

 Second, the TM may require extensive manual adjustments of its parameter settings to make it work satisfactorily for different images. In other words, raw images from different experiments or conditions cannot use the same parameter setting. Sometimes, manual adjustment is needed for different images in the same video (see Figure \ref{comporientation} for an example). This greatly hinders large-scale reliable studies of active nematics. Many factors contribute to this challenge,  such as: 1) active nematic materials can exhibit significant variations even under identical experimental conditions; 2)  different imaging instruments or settings can lead to different noise distributions; that is, different characteristics or statistical patterns of errors in the acquired images; 3) the behavior of nematic materials may not be consistent across the sample (e.g. due to spatially varying activity levels or boundary conditions, which may be according to design or arise from experimentally uncontrollable variations), resulting in defects with varying sizes and mean separation distances that require different configurations of window sizes. 

\subsection{Innovation and contribution}
\begin{figure*}
    \centering
    \includegraphics[width=0.6\linewidth]{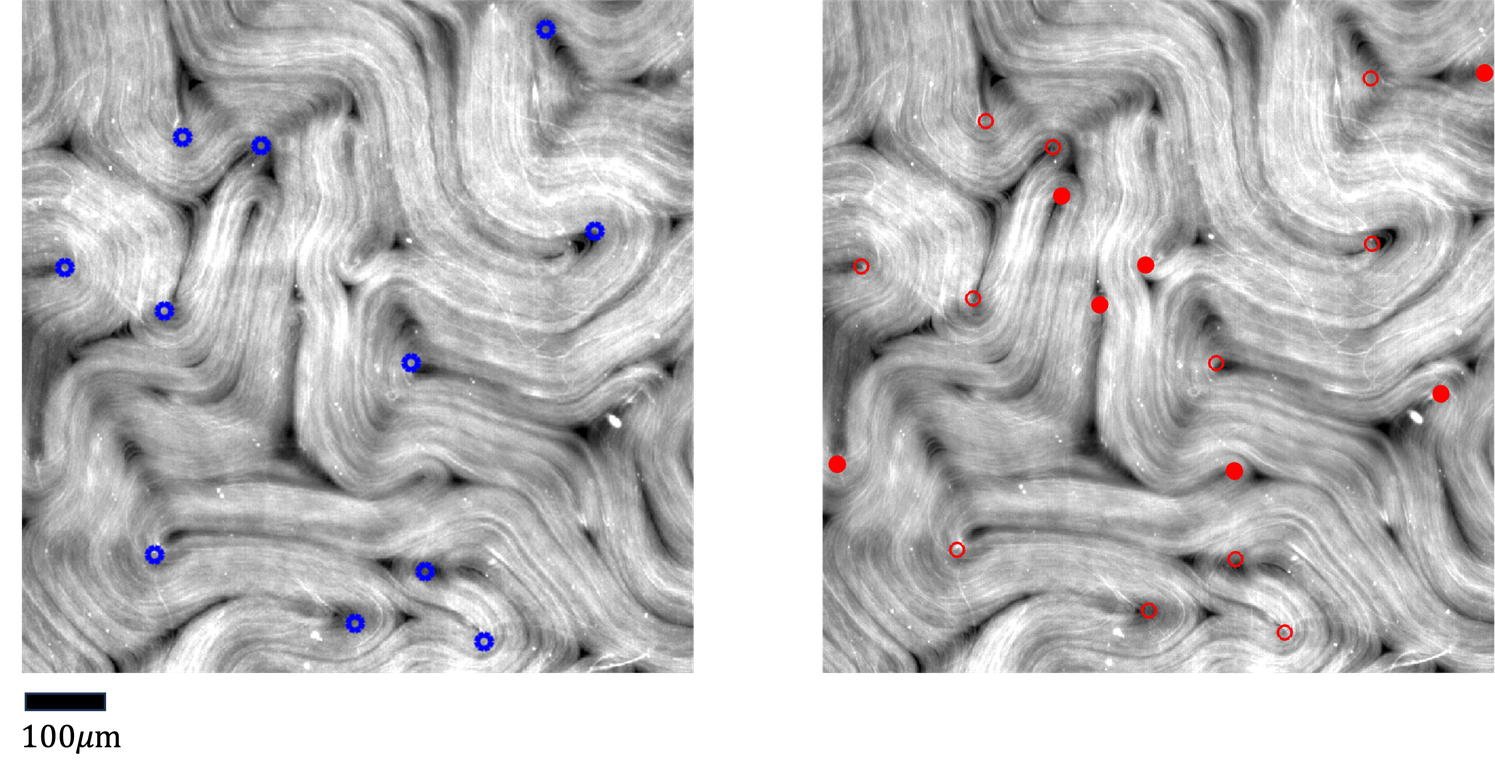}
    \caption{Comparison of defect detection by the  TM and our machine learning method. \textbf{Left}: Positive defects detected in the director field calculated by the TM. \textbf{Right}: Positive defects calculated by our machine learning method. Defects missed by the TM are highlighted in solid red dots. Empty red dots indicate the defects detected in both settings. Only detected positive defects are shown for clarity. The images are 600 by 600 pixels. The defect detection window in the TM is set to 11 and the defect threshold is set to 0.2. We note that the traditional method defect detection results may be slightly different if these parameters are tuned differently (Fig. \ref{fig:changeParameters} in Appendix A.6), but the parameters are tuned for each set of experiments so that all frames in a given experiment achieve acceptable results. Fig.~\ref{fig:changeParameters} (Appendix A.6) shows how these results depend on parameter settings.}
    \label{defect}
\end{figure*}

To tackle the above challenges, we developed a machine learning based technique to robustly and automatically extract director fields from raw images. Machine learning has already made significant impacts on active nematics \cite{cichos2020machine, rabault2017performing, hannel2018machine, munoz2020single}, and more applications are being explored. For example, we developed a machine learning technique for learning and predicting dynamics of active nematics \cite{zhou2021machine}, which however relies on the availability of director fields. Using our proposed machine learning approach, not only can we reliably and robustly calculate the director fields from raw experimental data directly, but we also enable many downstream tasks such as defect detection, defect tracking, and dynamic prediction.  These tasks are significantly simplified, compared to previous approaches, for raw data that does not have the director fields readily available. We benchmark our machine learning model against our in-house implementation of an image processing method (TM) that follows a similar approach to other existing methods\cite{tan2019topological, shi2013topological, blackwell2016microscopic, duclos2020topological, rezakhaniha2012experimental, ellis2018curvature}. This method takes a gray-scaled image as the input. After denoising the raw image by Gaussian Blur with standard deviation $\sigma$, it assumes that image intensities change most dramatically in the direction perpendicular to the flow of the microtubule bundles at each point. Thus, for each pixel, it calculates the intensity change matrix using convolution, which is then decomposed to obtain its eigenvalues.  The local orientation centered at a specific pixel is indicated by the eigenvector associated with the smallest eigenvalue. Once the $Q$-tensor is obtained, topological defects can be identified by calculating the winding number \cite{mermin1979topological} of each point. This is typically done by applying convolutional filters (e.g., Gaussian Kernels) of user-defined window size on each pixel on the director field obtained. More details of this method can be found in Appendix A.1. Compared with the TM, the director field information obtained by our method is more accurate, which leads to better defect detection results. Figure \ref{defect} shows an example of defects that are missed by the  TM. These defects can be recovered using our machine learning models.  We note that the TM defect detection results differ slightly when different parameter settings are chosen. Fig.~\ref{fig:changeParameters} (Appendix A.6) shows examples of changing the parameter settings for individual frames. However, the TM results are consistently less accurate than those obtained by our machine learning approach, and we have tuned the TM parameters for each dataset to achieve reliable results for all frames in that dataset. In addition to increased accuracy, a key advantage of the machine learning approach is that it requires less parameter tuning.

In summary, the main contributions of our work include the following:

\begin{itemize}
    \item We developed a robust machine learning approach to accurately extract director fields from raw microscopy images of active nematics. Our approach is able to handle raw images captured under various experimental settings (e.g., lighting, activity levels of active nematics, microscope lens, etc.). 
    \item  We used a masking strategy to correct some regions of director fields calculated by the PolScope that are inaccurate due to sub-optimal retardance images, such as over-/under-exposure, out-of-focus regions, etc. 
    \item We demonstrate promising downstream applications using the director fields generated by our model, including defect detection, defect orientation, defect tracking, etc.  
    \item To facilitate defect tracking as a downstream task, we utilized a matching algorithm across several consecutive frames and evaluated outcomes across frames. This approach enabled us to enhance tracking accuracy by filling in gaps for absent defects in a particular frame, while not in others.
\end{itemize}

\section{Methods}
We used the PolScope data (each sample is a pair of a retardance image and its corresponding director field) to train a deep learning model for extracting local orientation information from retardance images. Using the director fields extracted by our approach, we were able to improve defect detection and tracking.

\subsection{Improving training data}
\begin{figure}[!h]
    \centering
    \includegraphics[width=0.95\linewidth]{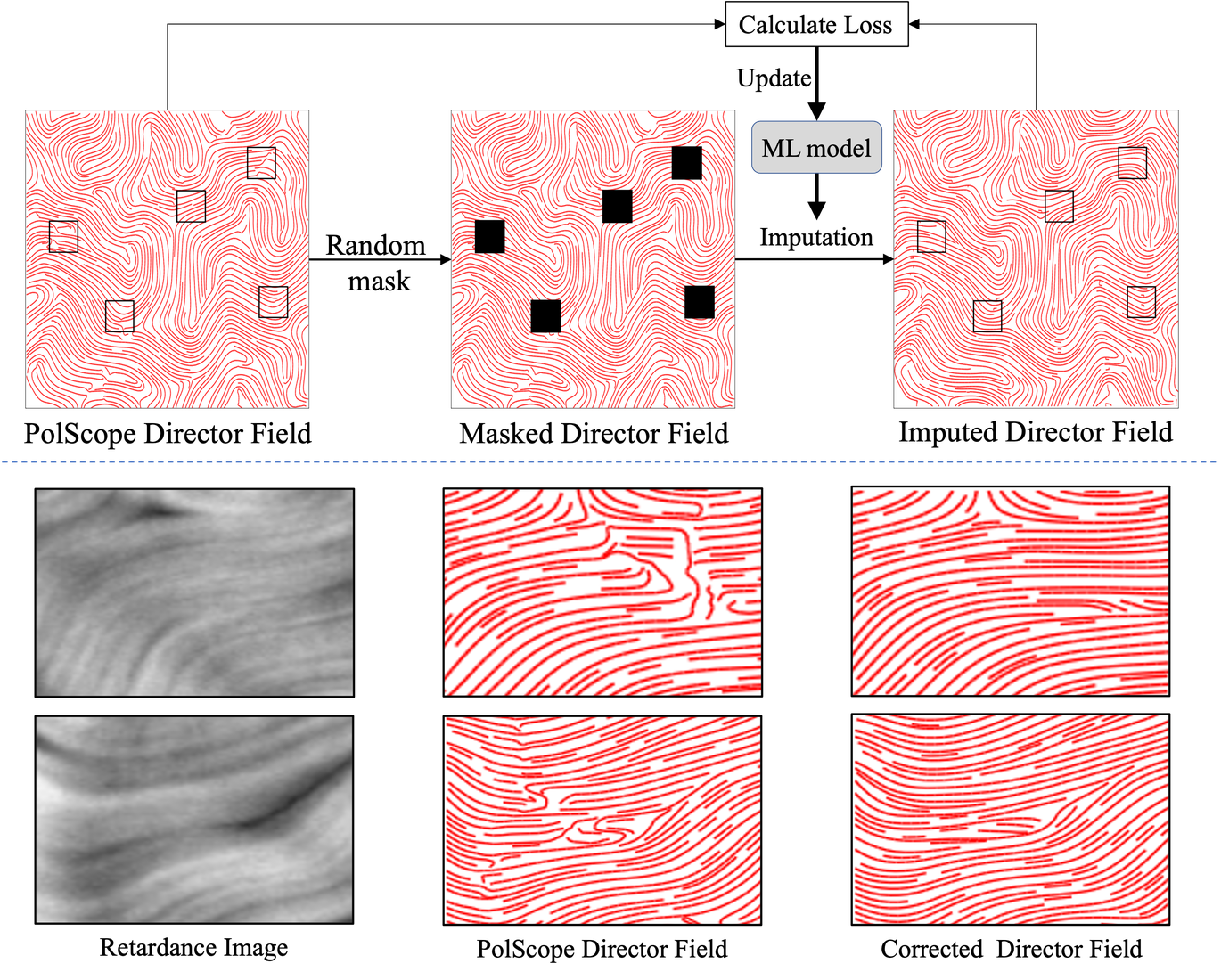}
    \caption{Correcting the PolScope director fields via Masked Image Modeling (MIM), which trains a machine learning model (the Swin Transformer\cite{liu2022video} in our case) to improve local orientation calculation. \textbf{Top:} In each iteration of the training process, a director field is sampled and several patches in it are randomly masked out (e.g., black patches in the middle image). The MIM machine learning model imputes the masked patches. The loss is calculated as the mean square error between the imputation results and the ground-truth (i.e., the corresponding patches in the input), and is used to update the parameters of the MIM machine learning model. \textbf{Bottom:} Two examples of local orientation correction are shown. The left, middle, and right columns are the retardance images, the  PolScope director fields, and the corrected director fields, respectively.}
    \label{model3}
\end{figure}

In most cases, PolScope works well in capturing the retardance images of active nematics and correctly calculating the corresponding director fields. However, PolScope occasionally produces sub-optimal local orientation calculations (examples in Figure \ref{fig:unsmooth}), which can hamper the efforts to build a machine learning model to extract director fields. Fortunately, such errors can be automatically corrected using machine learning. Specifically, we adopted the Masked Image Modeling (MIM) \cite{xie2022simmim} method (see Figure \ref{model3} Top) that randomly masks out patches in an image decided by a mask size and a masking ratio, and trained a machine learning model to reproduce the masked-out patches using the rest of the image.  We chose the mask size as 10x10 pixels and the masking ratio as 30\% of the whole image. To make this selection, we tried several different mask sizes (6x6, 10x10 and 20x20 pixels) and different masking ratios (20\%, 30\% and 50\%), and found the chosen setting produced the lowest Mean Absolute Error (MAE) of 0.0241. Once trained, the MIM model can be applied to improve director fields not used in training. Two examples are shown in Figure \ref{model3} Bottom to visualize the effects of this process.

\subsection{Director field extraction using machine learning}

\begin{figure*}[ht]
    \centering    \includegraphics[width=0.9\textwidth]{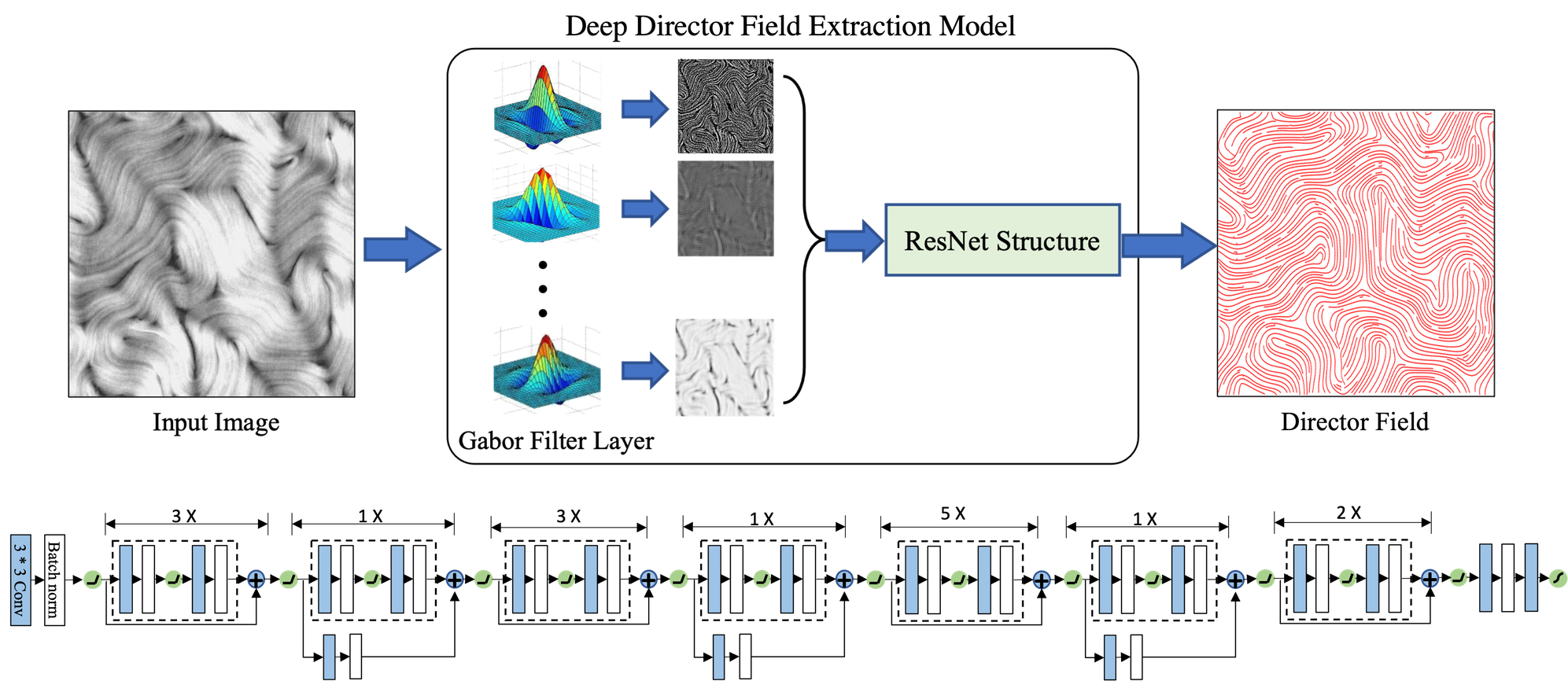}
    \caption{The deep learning model for extracting director fields from PolScope retardance images. \textbf{Top}: Complete model pipeline. \textbf{Bottom}: Details of the ResNet Structure in \textbf{Top}. Blue and white boxes indicate convolution layers and batch normalization operations. A ReLU activation function \cite{maas2013rectifier} is used except for the output layer, which uses a sigmoid activation function. In total, 16 ResNet blocks are used. }
    \label{model12}
\end{figure*}

Using the corrected director field as the target output, we can train our machine learning model to learn the proper mapping between the retardance image to its corresponding director field. Figure \ref{model12} illustrates the architecture of our machine learning model, which consists of a Gabor filter \cite{fogel1989gabor, bianconi2007evaluation} layer followed by a ResNet structure \cite{He_2016_CVPR}. Gabor filters are a set of filters widely used for texture analysis. Each filter can be tuned to detect spatially local signals with certain frequency components in a certain direction at a certain scale. One can use an array of Gabor filters to help perform multi-scale analysis of spatial patterns in active nematics, such as defects, vortices, or other patterns. More technical details about Gabor filters are provided in Appendix A.2. In our model, we used two sets of Gabor Filters, the first set has a smaller wavelength that aims to detect higher spatial frequencies and finer details, while the second set has a larger wavelength, aiming to capture global features. For each set of filters, 64 filters are used, each with a different combination of 8 orientation values and 8 scale values. For the ResNet, 16 blocks are used, as shown in Figure \ref{model12}. Among the 16 ResNet blocks, 4 different channel dimensions are used: 64, 128, 256, and 512. More detailed configurations can be found in Appendix A.3.

\subsection{Defect analysis}
\begin{figure*}[ht]
    \centering
    \includegraphics[width=0.7\linewidth]{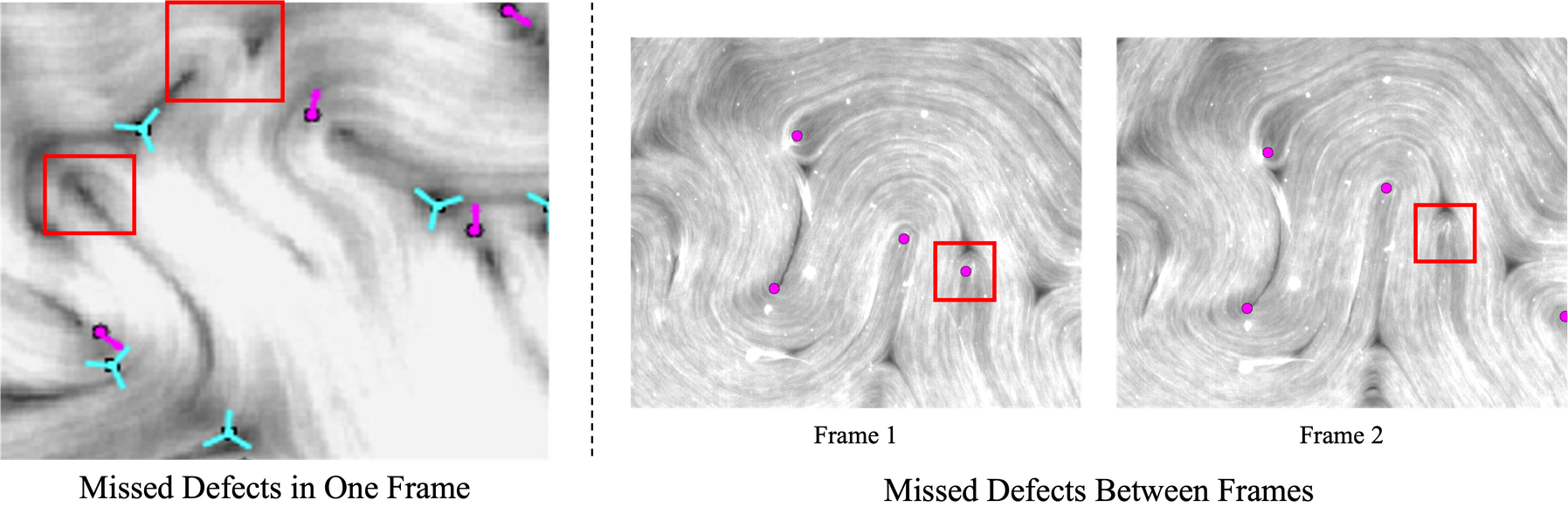}
    \caption{Example situations that lead to inaccurate defect tracking. \textbf{Left: Challenging raw images.} A raw image with a high density of defects while the light exposure is uneven. The winding number calculation from director fields generated by the machine learning model can be less accurate in comparison to regions with even exposure.  Red arrows show positive defects and blue tricuspoids show negative defects. Red rectangles mark places where defects are missed. \textbf{Right: Inconsistency between frames.} As the active nematic system evolves, some defects can be detected in one frame but may be missed in another.  This could lead to inaccurate conclusions of defect statistics such as their rates of annihilation and proliferation. When tracking defects in a sequence, the missed defects can be interpolated by our tracking algorithm. Only positive defects are marked by red dots for clarity.}
    \label{defect_challenge}
\end{figure*}

One can detect defects by computing the winding numbers of individual points in the director fields \cite{mermin1979topological} by counting the local orientation changes centralized at each point within a predefined window size. Nevertheless,  the director fields are discretized at the pixel level, which can lead to inaccurate winding number calculations. Some defects may not be detected even from more accurate director fields such as those calculated by PolScope (see Figure \ref{polscope_defect} in Appendix A.6) and our machine learning approach (see Figure \ref{defect_challenge}, Left), due to various factors. These include over-exposed regions in the raw image, leading to an over-smoothed director field; or regions with high local defect densities, where the distance between the defects is smaller than the window size when calculating the winding number. 
The director fields calculated by our method significantly improve defect detection results (see Figure \ref{flore_compare} in the Experiments section). However, in a non-equilibrium system of active nematics where speed and density change, it is possible that a defect is detected in one frame but missed in the next (see Figure \ref{defect_challenge} Right). This can cause the defect tracking algorithm to lose track of detects in consecutive frames and result in inaccurate measurement of defect annihilation and proliferation rates. 

To tackle this problem, we match defects detected in three consecutive frames (at time points $t-1$, $t$, and $t+1$) using a graduated assignment algorithm \cite{gold1996graduated}, and use the matching results to recover missing defects in the $t$-th frame. The coordinates of the rescued defects are interpolated using the coordinates of their matched defects at time $t-1$ and $t+1$.  This allows for more accurate defect detection and tracking over time, see Figure \ref{lowatp}, Figure \ref{highatp}, and Figure \ref{defect_threshold}. Please refer to Section 3.3 and 3.4 for more results and discussions.

Finally, we trained a model with four ResNet blocks (32, 64, 128, and 256 channels, respectively) to estimate the orientation of a defect. The input to the model is a 64 $\times$ 64 patch centered at a defect, and the output is the angle of the defect. The model error is also measured by the Mean Absolute Error (MAE) between the estimated angles and the manually labeled ground-truth.

\section{Experiments}
In this section, we will illustrate our training process and model performance. In total, 48,000 retardance and director field images are used for training and testing the machine learning model. Specifically, we randomly selected 40,000 image pairs to train the neural network, and the remaining 8,000 images for performance testing. We present results obtained by applying the model to the test set of data. We chose the PolScope dataset to train the model because it is the only instrument that can produce retardance and director field images at the same time. The director field images are first corrected using the masking strategy, then used as the training targets for the machine learning model. The retardance images are inputs to our machine learning model. The predicted director field is compared with the corrected director field and the learning goal is to achieve maximum similarity. Since the orientations of microtubule bundles have head-tail symmetry, we translate the orienation to the $Q$-tensor and use the Mean Squared Error (MSE) to measure the learning error as follows. 
$$
    Q_{xx} = \cos^2(q_i), \quad
    Q_{xy} = \cos(q_i)\sin(q_i), $$
    $$
    \Loss = \frac{1}{D} \sum_{i=1}^{D}\left[(Q_{xx}^\text{predict}- Q_{xx}^\text{target})^2 + (Q_{xy}^\text{predict} - Q_{xy}^\text{target})^2\right]
$$
where $q$ is the director field orientation angle at each pixel in the input images. 

\subsection{Extract director fields from retardance images}

\begin{figure*}[!ht] 
    \centering
    \includegraphics[width=0.7\textwidth]{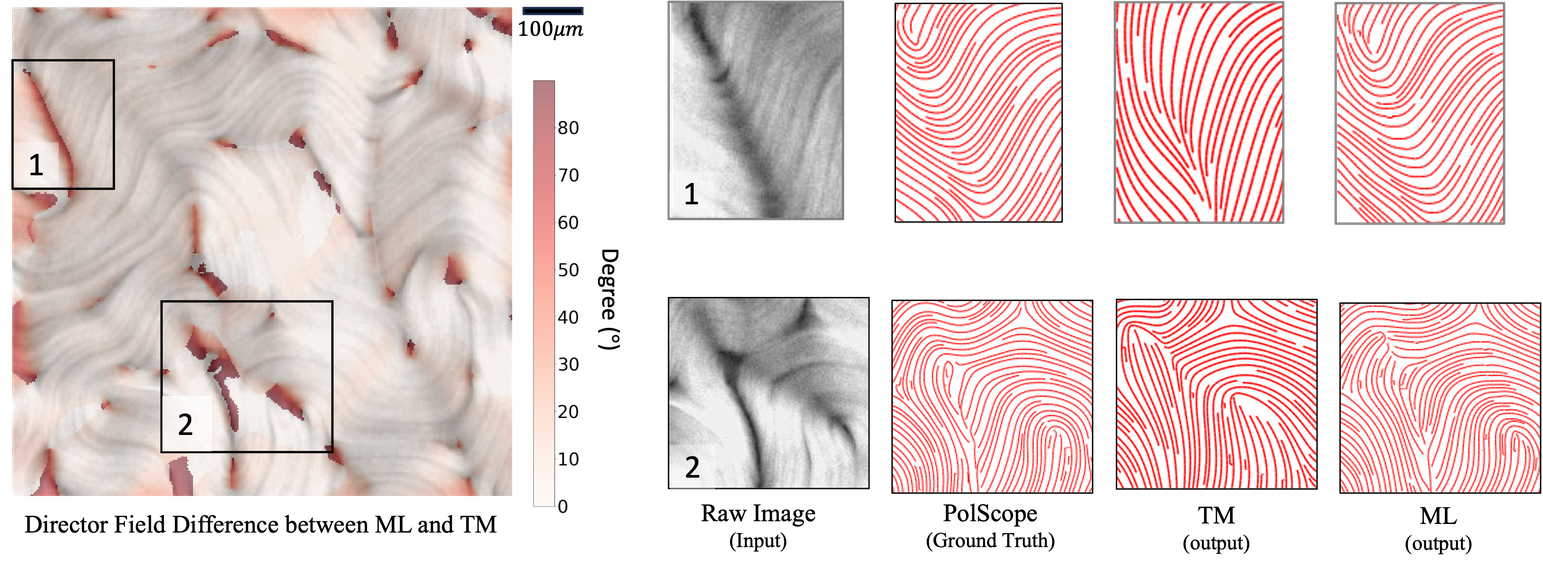}
    \caption{Comparison of the director fields extracted from PolScope retardance images by the traditional image analysis method (TM) and our machine learning (ML) method. \textbf{Left:} Heatmap showing the difference between the director fields calculated by TM and ML, overlaid on the retardance image. Darker colors indicate bigger discrepencies in local orientation measured by degree. \textbf{Right:} Zoom-in views of two regions (1 \& 2) marked in the Left showing director fields calculated by PolScope  (which are used as the ground truth),  TM, and ML. $\sigma$ is set to 13.}
    \label{polscope}
\end{figure*}

Figure \ref{polscope} illustrates the advantage of our machine learning model over the TM on the PolScope retardance images.  Using the polished director fields calculated by PolScope as the ground-truth, our trained machine learning model extracts more accurate director fields compared to the TM. We tested our model on 120 randomly selected retardance images excluded from the training set. Using the corresponding director fields calculated by PolScope as the ground-truth, we calculated the error of our model and the TM method. Our approach achieves a mean absolute error (MAE) of 8.8 degrees per location with a standard deviation of 1.46. As a comparison, the MAE of the TM approach is 18.1 per location with a standard deviation of 2.6.

\subsection{Generalize to fluorescence images}
 The proposed machine learning approach is trained using PolScope retardance images and calculated director fields. However, once trained, it can be applied to raw images produced at other magnifications, by other microscope instruments, or under different experimental settings (e.g., activity levels or boundary conditions). There are various methods to control the behavior of active nematics, such as changing the concentration of ATP and motor proteins, applying hard boundaries, etc \cite{keber2014topology,opathalage2019self,thijssen2021submersed,zhao2020stability,hardouin2019reconfigurable,lemma2019statistical}. However, these methods control activity with only limited precision, and do not enable spatially varying activity profiles. To enable greater control over activity and the ability to change activity in space and time, researchers have developed light-sensitive motor proteins  \cite{zhang2021spatiotemporal, lemma2022spatiotemporal, ross2019controlling}, with which the active nematic's activity and dynamics can be regulated by varying the spatiotemporal patterns of applied light intensity. 

\begin{figure}[h]
    \centering
    \includegraphics[width=0.95\linewidth]{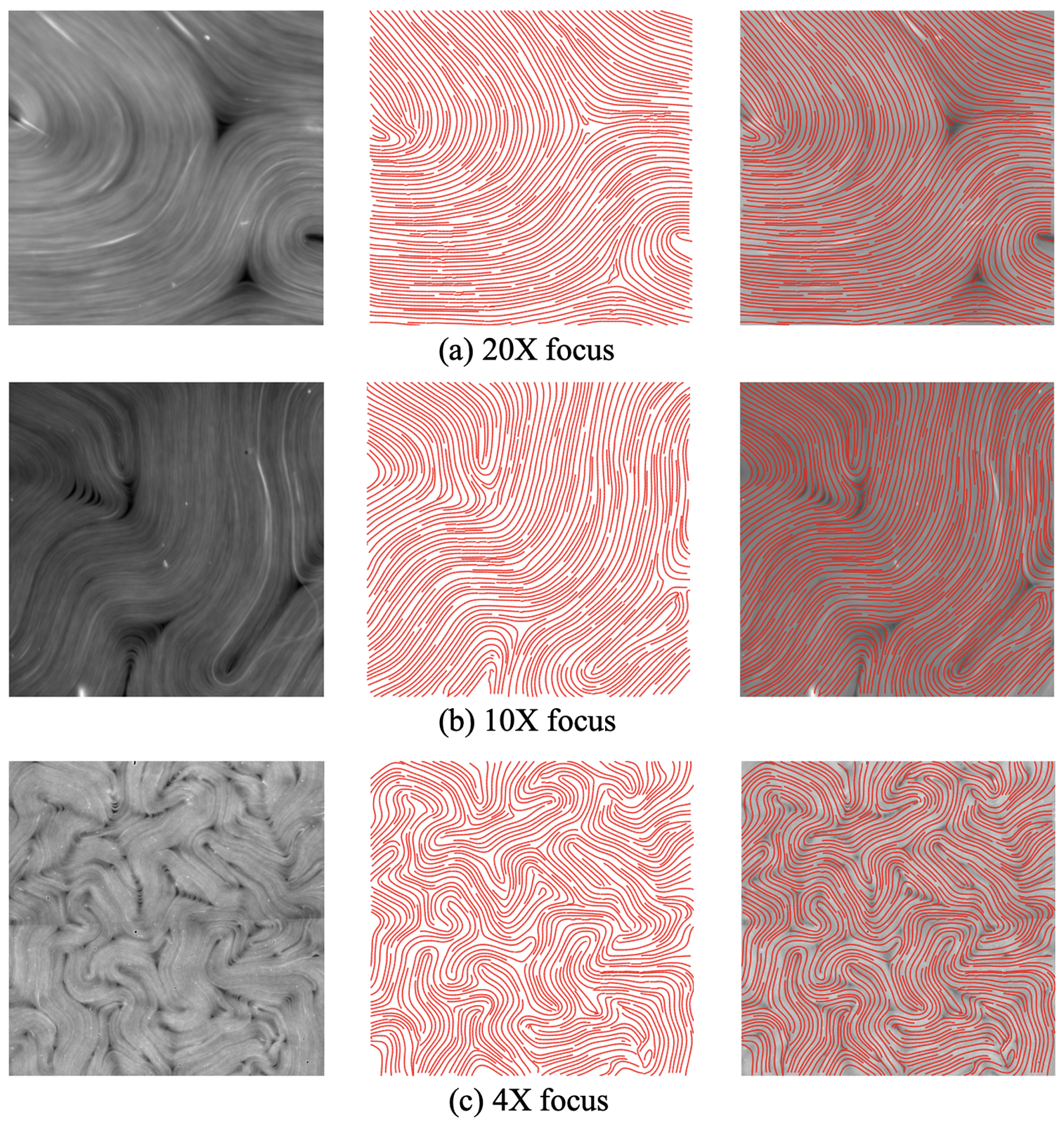}
    \caption{Exemplar performance of the machine learning (ML) model on fluorescence images with different microscope magnification settings. The left, middle, and right columns are the input raw images, director fields extracted by our ML model, and director fields superimposed on raw images, respectively. The model was trained on the retardance and director field data captured at 4x. It generalizes well to fluorescence images captured with microscope settings that were not used to produce training data.  Within the RAM memory constraint, our model can calculate the director field for varying input sizes. Here, all images are cropped to 512 by 512 pixels for fair comparison.}
    \label{focus}
\end{figure}

To create an active nematic, we combined microtubules, kinesin motors, ATP, and bundling agents, and sediment the mixture on an oil-water interface. In the light-activated active nematic, the kinesin motors are equipped with light-sensitive domains, namely ``iLID''and ``micro''. When exposed to blue light, these domains bind together rigidly, creating a light-sensitive cluster that generates relative sliding for anti-parallel microtubules. To illuminate the sample, we used a Digital Light Processor (DLP) projector that projected 460 nm wavelength light from above onto the sample and we took data using a fluorescent microscope. However, the PolScope setup, which involves a universal compensator and an analyzer for circularly polarized light, limits access to the stage and does not allow us to use the DLP mounted to the stage for light-activated experiments. Additionally, data acquisition with PolScope requires taking 5 images in different polarization settings to calculate the retardance image and its director field \cite{shribak2003techniques}, which limits the frame rate for taking data.  For example, with a camera exposure time of 500 ms, it takes 670 ms to capture a single image and roughly 3.4 sec for five images. The impact of these exposure times on the observations becomes noticeable when the nematic speed is high, as the motion between frames is difficult to document, and consequently, tracking defects becomes less precise. While the PolScope microscope has some advantages, such as calculating the director field without the need to label the proteins, it is not suitable for all experiments, such as the light-activated active nematic. Hence, it is desirable that our director field extraction model, which was trained on PolScope retardance images, can generalize well to other raw images.  Our proposed machine learning model performs well on fluorescence images of active nematics with different microscope magnifications (see Figure \ref{focus}) in an unconfined 2-dimensional experimental setting. Moreover, our model also generalize well to confined containers with different boundary conditions (see Figure \ref{annulus} in Appendix A.6, and videos in the supplementary material). The authors highlight that even in the presence of unbound microtubules floating along the z-axis of the 3D container, which is not aligned with the local orientation within the 2D nematic system under investigation, the machine learning algorithm adeptly filters out these elements, ensuring reliable calculation of the director field.

\subsection{Defect detection}
The defect detection results using the director fields extracted by our approach are substantially better than those obtained using the director fields extracted by the TM (details in Appendix A.1). Figure \ref{flore_compare} visualizes the defect detection results from raw images in one experiment using fluorescence microscope, which was not used to train our director field extraction model.

\begin{figure*}[!ht]
    \centering
    \includegraphics[width=0.6\textwidth]{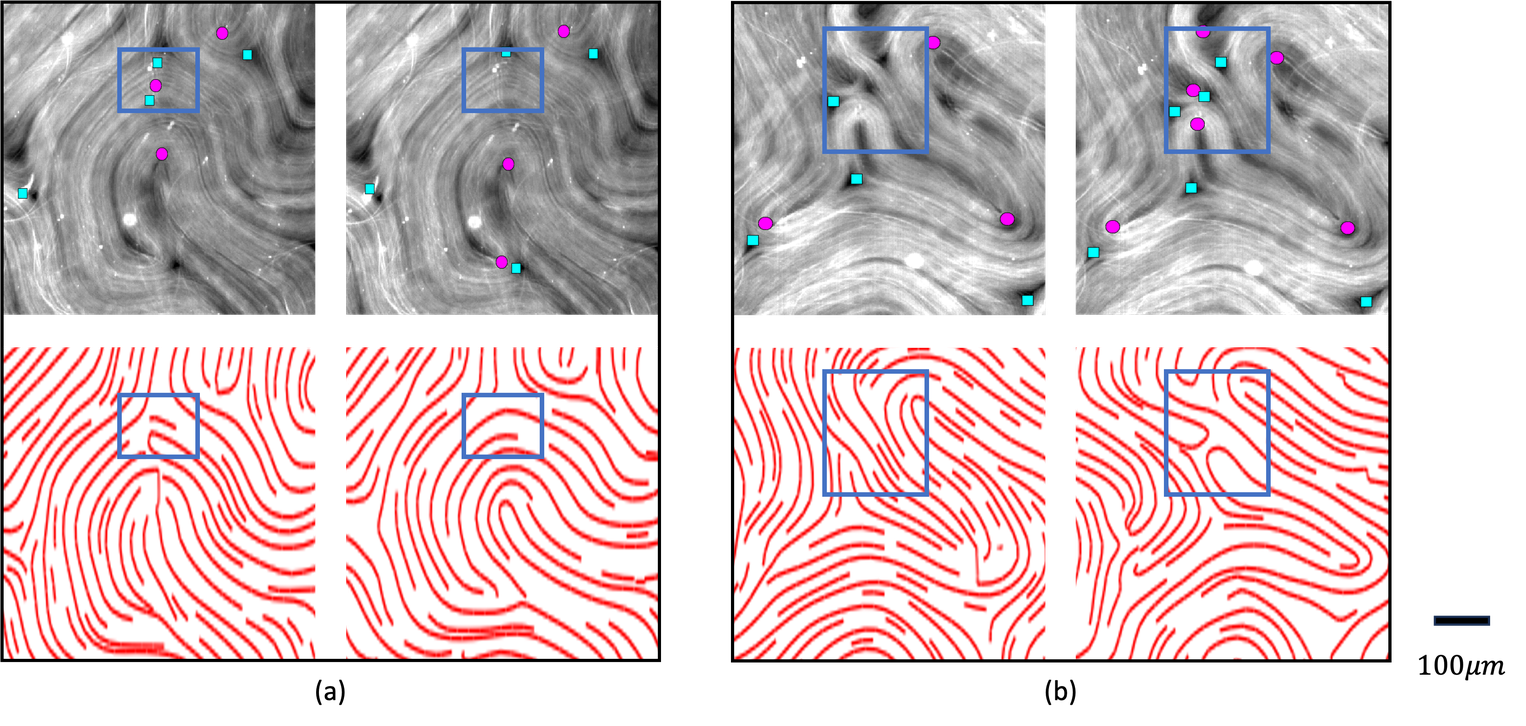}
    \caption{Example results of our approach to improve defect detection results on fluorescence images. \textbf{(a)} The boxes highlight the regions where the TM \textbf{(left)} produces false positives, while our method \textbf{(right)} produces correct results. \textbf{(b)} The boxes highlight one noisy region where the TM \textbf{(left)} fails to detect the positive and negative defect pair. Our method \textbf{(right)} is less sensitive to such noise and succeeds in detecting the defect pair. $\sigma$ is set to 11.}
    \label{flore_compare}
\end{figure*}

\begin{figure}[!ht]
    \centering
    \includegraphics[width=0.7\linewidth]{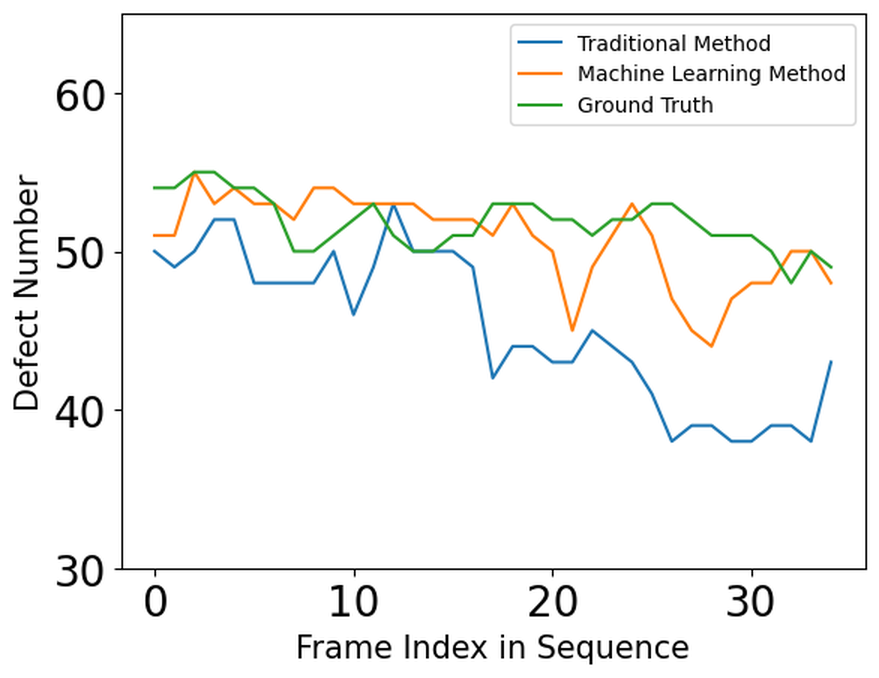}
    \caption{Comparison of defect detection performance by the TM and machine learning  ML methods on fluorescence image sequences captured under low lighting intensity  ($0.13$ mW/cm$^2$).  The interval between frames is 7 seconds. The green curves indicate the ground-truth, which was manually annotated. The blue and red curves are the defect detection results using the director fields calculated by the TM and our ML method respectively. On average, the TM produces 6.75 detection errors per frame whereas the ML method produces 2.25 detection errors per frame. For the TM, the window size is set to 9 and the defect threshold is set to 0.2.}
    \label{lowatp}
\end{figure}

\begin{figure}[!ht]
    \centering
    \includegraphics[width=0.9\linewidth]{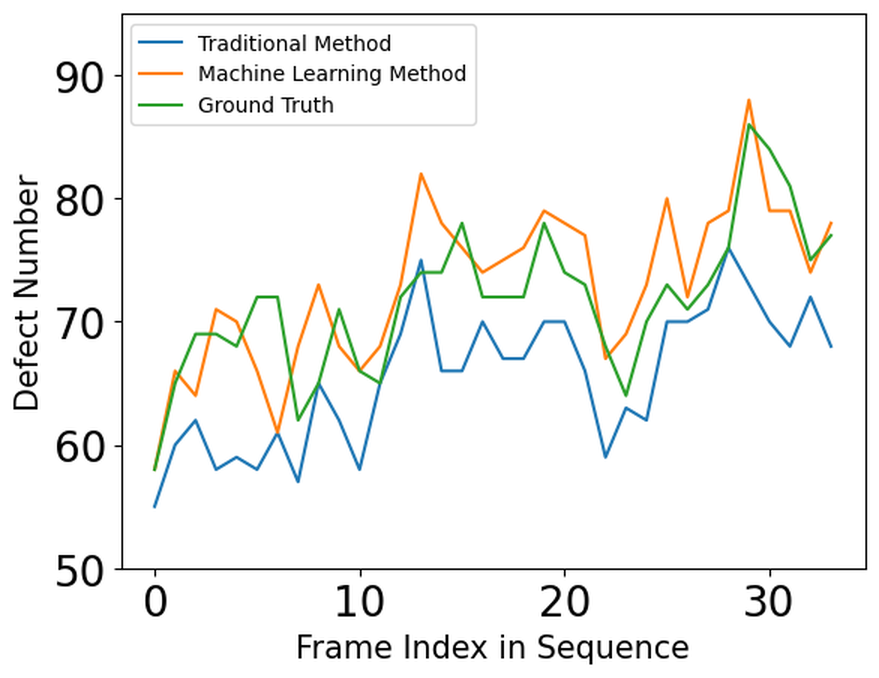}
    \caption{Comparison of defect detection performance by the TM and machine learning  ML methods on fluorescence image sequences captured under high lighting intensity ($2.01$ mW/cm$^2$). On average, the TM produces 6.26 detection errors per frame whereas our ML method produces 3.41 detection errors per frame.  For the TM, the window size is set to 11 and the defect threshold is set to 0.3.}
    \label{highatp}
\end{figure}

\begin{figure}[!ht]
    \centering
    \includegraphics[width=0.9\linewidth]{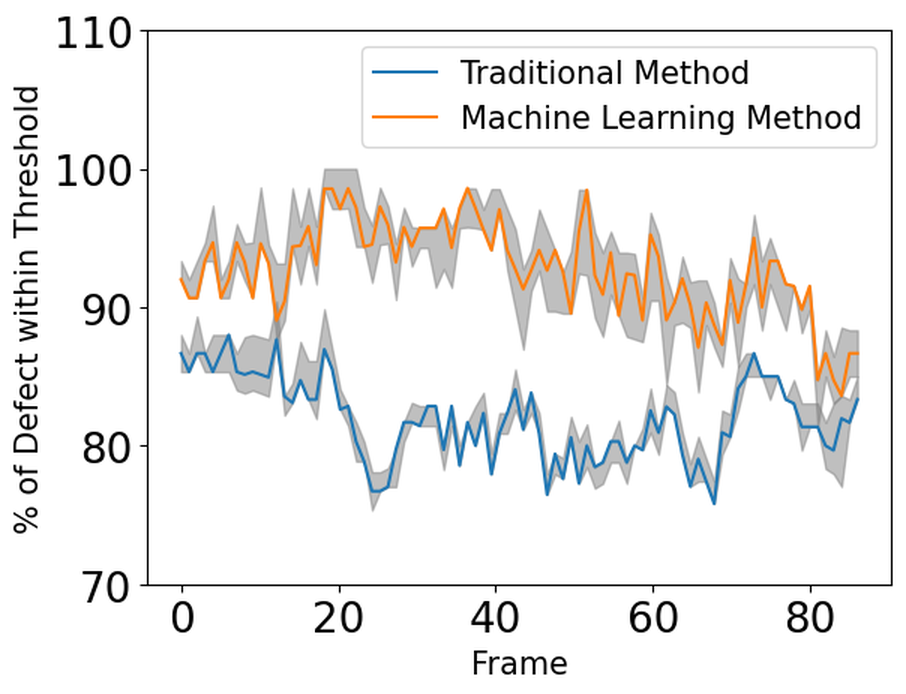}
    \caption{Precision of defect detection by TM and ML methods. The percentage of precise defect detections is shown for each method for a raw image size of $1296 \times 1280$. A defect detection is considered precise if the detected defect is within a  threshold distance from the manually labeled defect.  \textbf{Center Line}: threshold of 40 pixels. \textbf{Lower Boundary}: threshold of 30 pixels. \textbf{Upper Boundary}: threshold of 50 pixels. The defect threshold is set to 0.2.}
    \label{defect_threshold}
\end{figure}

To evaluate the defect detection results, we manually labeled defects in 40 fluorescence images  (not used in model training) from one experiment as the ground-truth. Fluorescence microscopy enables a light-responsive system that allows researchers to combine a considerable quantity of ATP with microtubules and regulate its behavior using light intensity. When the light intensity is increased, a larger number of motor proteins and microtubules become attached, leading to increased system activity and ultimately resulting in more defects. Figure \ref{lowatp} and \ref{highatp} show the performance of our approach in both low and high light intensity experiments. Note that the high-intensity sample (Figure \ref{highatp}) has not reached steady-state, and thus the number of defects is increasing over time. Using the director fields extracted by our machine learning model, the numbers of defects closely track the hand-labeled ground-truth.  Moreover, we closely observed the defect location generated by both methods. Compared with the manually annotated location, if the detected location is within a 40-pixel radius,  which is roughly half of the averaged defect size (100um, 77 pixels), it is considered a precise detection.  The size of the raw image is $1296 \times 1280$. Figure \ref{defect_threshold} compares the percentage of precise detections using both methods. The director field extracted by our machine learning model enables better defect detection results (both defect detection rate and location).

\subsection{Defect tracking}

\begin{figure}[!h]
    \centering
    \includegraphics[width=0.9\linewidth]{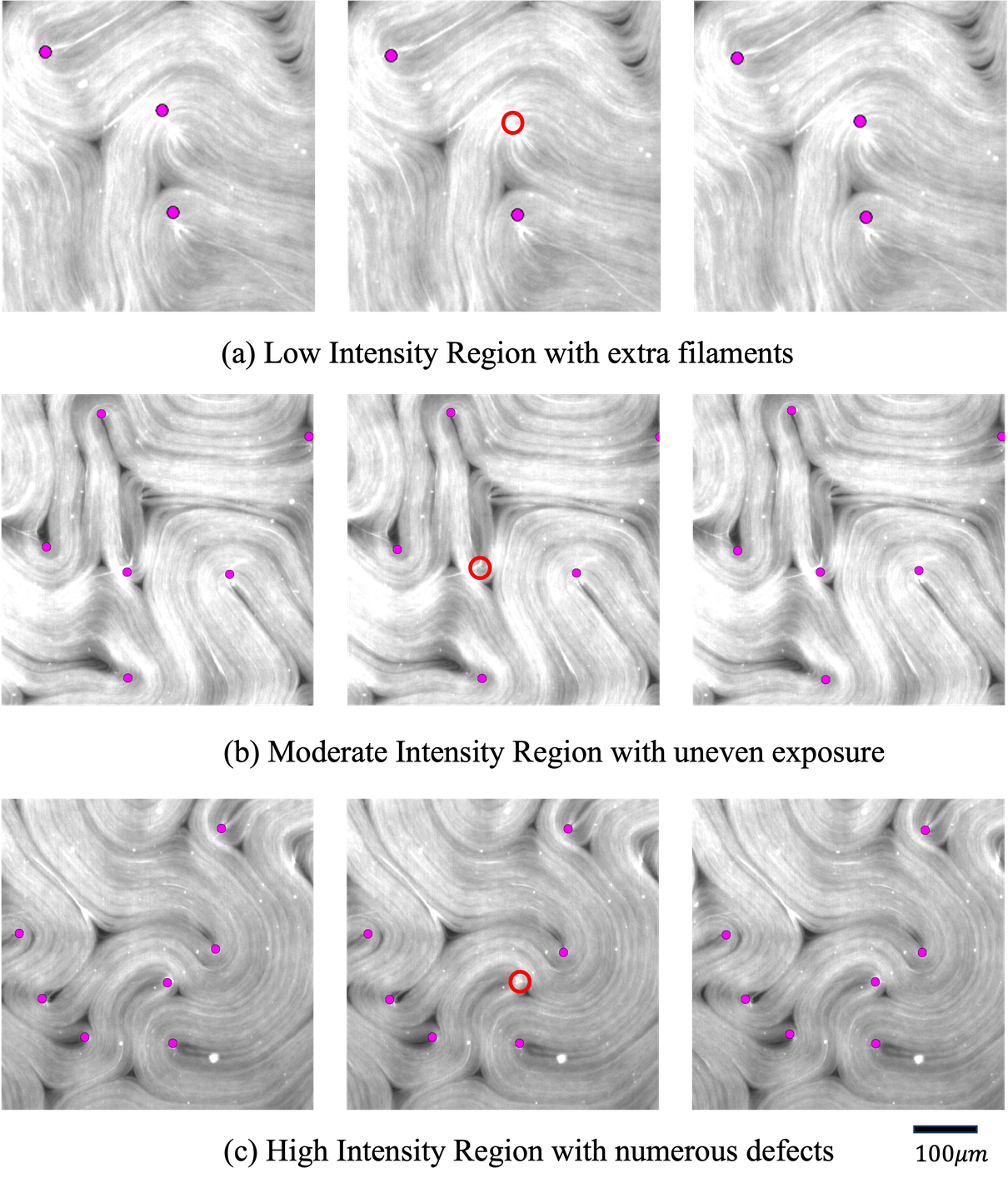}
    \caption{Examples of recovering missed defects. Fluorescence images were used, and the examples are from different regions of the sample, where defect density and lighting conditions vary. Each row shows three consecutive frames. The missing positive defects (red circle) in the middle frame are interpolated using their counterparts in the preceding and succeeding frames. Only positive defects are shown for clarity.}
    \label{tracking_intp}
\end{figure}

Once defects are detected, we can track them over time. This can be achieved by matching the locations of defects across different frames using the graduated assignment algorithm \cite{gold1996graduated}.  This article proposed a fast and accurate graph matching algorithm using nonlinear optimization and graduated non-convexity to avoid poor local minima. This algorithm allows matching between graphs of different node numbers. This fits well with our system because it enables handling the spontaneous birth and annihilation of defects. However, it is possible that one defect is detected in one frame,  but disappears in the next frame. This could be due to defect annihilation or a mistake in the defect detection process. To deal with the latter, we apply the matching algorithm to three consecutive frames. Using the matching results, we can interpolate the locations of the defects that are missed in the middle frame but are detected in the preceding and succeeding frames.  Figure \ref{tracking_intp} illustrates the tracking results on three image sequences captured under different lighting conditions.


\subsection{Defect angle estimation}
\begin{figure}[!h]
    \centering
    \includegraphics[width=0.9\linewidth]{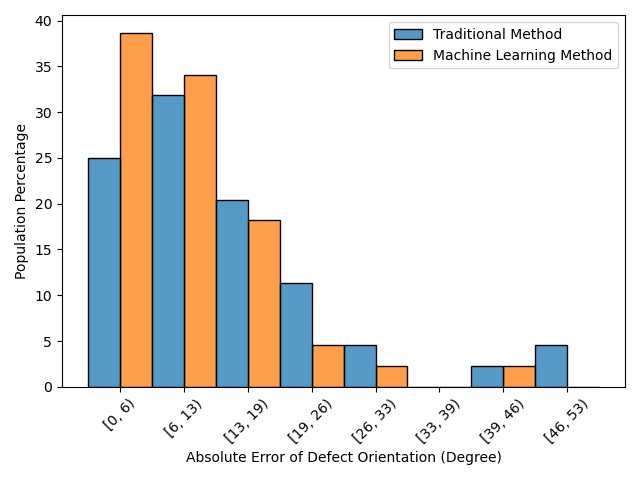}
    \caption{Error distribution of the estimated detect orientations in absolute angle differences. Compared with the manually labeled orientation for each defect, the machine learning model achieved an average error of $9.86^\circ$, whereas the TM error is $14.33^\circ$.}
    \label{angles_compare}
\end{figure}

One natural extension from the foundation of an accurate director field is to identify the microtubule angles at the defect locations, which may provide insights into the dynamics of the active nematic system. This can be conveniently achieved using a similar CNN network introduced in Section 2.2 and Section 2.3. The image processing method (TM) can also derive approximate defect orientations (details are in Appendix A.1). To obtain accurate angles at defect locations, we manually labeled 800 defects using the Python package ``PyQt5'', out of which, 390 were negative and 410 were positive defects. We ran our model and the TM on the same images and compared the angle differences with the ground-truth. Among these defects, our model and the TM resulted in average errors (defined as the difference in detected angle from the ground-truth angle) of $9.86^\circ$ and $14.33^\circ$ respectively. The comparison results are summarized in Figure \ref{angles_compare}.

\section{Conclusions and Discussions}
We present a machine learning based approach for automatically extracting director fields from images of active nematics, which enables downstream analyses, such as topological defect detection and tracking, defect orientation estimation, and related characteristics. Our approach is less sensitive to noise and can generalize to images captured by settings (e.g., microscope instruments, lighting conditions, and camera focus) not encountered in the training phase. Comparison of results from applying our method and the traditional method (TM) to experimental data demonstrates the superior performance of the machine learning approach.  Despite the advances relative to traditional image processing methods, we note that our method does not entirely eliminate noise, particularly in suboptimal regions such as overexposed areas or those containing additional microtubule filaments above the plane of the active nematics. Furthermore, as our model is trained on PolScope images, its performance may vary slightly when applied to raw images from other microscopy techniques due to differences in noise distributions. Nevertheless, machine learning models have consistently demonstrated their potential in the realm of Soft Matter research, providing precise and automated solutions for diverse tasks, from dynamic predictions \cite{zhou2021machine} to defect analyses \cite{minor2020end, chowdhury2023topological}. Our algorithm can be further improved by training on other types of microscopy images.

 By employing the suggested machine learning method, we can obtain director fields with greater accuracy. This enhances the reliability of detecting and pinpointing defects. Such improved data is vital for addressing scientific questions regarding the time taken for a system to reach a steady state, the spatial distribution of defects, and their rates of annihilation and proliferation. In the future, we plan to use the calculated director field to do dynamics forecasting. In other words, given a short time sequence of images from an active nematics experiment, we aim to predict the subsequent time evolution of the sample.  The more accurate calculation of director fields and detection of defects should lead to improvements over current algorithms \cite{zhou2021machine,colen2021machine}.
 Forecasting is the first step toward being able to control the behaviors of an active nematics system, by actuating with light or other external control inputs. Similarly, we hope to use this approach to derive statistics of the system's characteristics and link them with the dynamics of active nematic systems. More broadly, our approach could be generalized to other active matter and soft matter systems.

\clearpage
\appendix
\section{Appendix}

\subsection{Traditional image processing method used}
The traditional image processing method used as a benchmark in this paper is from \href{https://github.com/wearefor/qcon_nematicdefectfinder}{Michael M. Norton}. The steps are below:
\begin{itemize}
    \item Calculate the intensity change of a given image.
    \begin{itemize}
        \item Apply a Gaussian filter:
            $$G(x, y) = \frac{1}{2\pi \sigma^2} \exp{(-\frac{x^2 + y^2}{2\sigma^2})} * I(x, y),$$
            where $G(x, y)$ is the smoothed image, $\sigma$ is the standard deviation of the Gaussian filter, $I(x, y)$ is the original image, and $*$ is the convolutional operator.
        \item Compute the gradient:
            $$\nabla G(x,y) = \left( \nabla_x G,\nabla_y G\right)\text{,} \text{where} $$
            
            $$\nabla_x G(x, y) = \frac{\partial G}{\partial x}(x, y) \text{and} \nabla_y G(x, y) = \frac{\partial G}{\partial y}(x, y),$$ 
            
            
        \item Compute the gradient magnitude:
            $$\|\nabla G(x, y)\| = \sqrt{\left(\nabla_x G(x,y)\right)^2 + \left(\nabla_y G(x,y)\right)^2}.$$
        Optionally, compute the gradient direction:
            $$\theta(x, y) = \atantwo(\nabla_y G(x, y), \nabla_x G(x, y)),$$
            where $\theta$ is the angle of the gradient vector with respect to the x-axis, and  $\atantwo$ is the four-quadrant inverse tangent that returns values in the closed interval [$ - \pi$, $\pi$].
    \end{itemize}
    \item Derive orientation.
    \begin{itemize}
        \item Since the microtubule bundles have head-tail symmetry, we can use the rank-2 gradient tensor:
        $$H(x, y) = $$
        
        $$\begin{bmatrix}
            (\nabla_x G(x,y))^2 & \nabla_x G(x,y) \nabla_y G(x, y) \\
            \nabla_x G(x,y)\nabla_y G(x, y) & (\nabla_y G(x, y))^2
        \end{bmatrix}.
        $$
        \item Compute the eigenvalues of the gradient tensor:
            $$\lambda_1(x, y), \lambda_2(x, y) = $$
            $$\frac{1}{2} \left[ \trace(H(x, y)) \pm \sqrt{\trace(H(x, y))^2 - 4 \det(H(x, y))}\right],$$
            where $\trace(H(x, y))$
            
            is the trace of $H(x, y)$ and $\det(H(x, y))$ is the determinant of $H(x, y)$.
        \item The eigenvalues $\lambda_1$ and $\lambda_2$ represent the strength of the gradient at each pixel in the direction of the two corresponding eigenvectors. The director field should be the smaller of the two eigenvalues \cite{blackwell2016microscopic}. 
    \end{itemize}
\end{itemize}

After the director field is calculated, topological defects can be identified from regions of rapid rotation of the director field \cite{shi2013topological, tan2019topological}. The winding number is used to measure the change. The winding number is a mathematical concept that describes the number of times a curve, typically a closed curve, wraps around a given point in a plane.  In the context of liquid crystals and active nematics, the winding number is a topological concept that captures the number of times the director rotates around a closed loop in the director field. More formally, it quantifies how many times the director completes a full rotation (360 degrees) when one traverses a closed curve in the plane of the field. It is a useful tool in topology and geometry for characterizing the behavior of curves and shapes. To calculate the winding number using the director field, we can choose a point in the space and numerically count the local orientation change around this point. Applying a threshold, the topological defect can be detected.

 To deduce the intrinsic orientation of a defect, a circular window is defined around every detected defect location. For every pixel along this window, the local orientation of the director field is noted. Then, the angle originating from this pixel to the defect's center is computed. Next, we compute the scalar product of these two angles. Then, the pixel yielding the maximum scalar product value indicates the direction in the director field most closely aligned with a line radiating from the defect's core. We use the orientation of the director field in this specific pixel as a proxy to infer the intrinsic orientation of the defect.

\subsection{Details of the Gabor filter}

A Gabor filter \cite{fogel1989gabor} is a type of linear filter used in image processing that is designed to respond to specific frequencies or orientations in an image. In the context of Polscope images of active nematics, Gabor filters can be used to enhance the visibility of the nematic texture and reduce the effects of noise. Eq.~\eqref{Gabor} gives the equation of the Gabor filter. It is composed of two components: a complex sinusoidal wave and a Gaussian kernel. The Gaussian kernel allows the filter to respond to features of varying sizes, while the sinusoidal wave allows the filter to respond to features of varying orientations. By computing the Gabor response at multiple scales and orientations, features such as texture, edges, and corners can be identified and used to segment and classify the image.


\begin{align}
    \label{Gabor} 
    g_{\lambda,\theta, \phi, \sigma, \gamma}(x, y) = \exp{(-\frac{x'^2+\lambda^2 y'^2}{2 \sigma ^2})} \cos( \frac{2 \pi x'}{\lambda} + \phi)
\end{align}

where 
\begin{align*}
    x' = x \cos \theta + y \sin \theta, \quad   y' = -x \sin \theta + y \cos \theta.
\end{align*}

Here, $x$ and $y$ are the coordinates of image pixels, $\sigma$ is the width of the Gaussian kernel, $\theta$ is the angle of the wave function, $\phi$ is the phase, and $\lambda$ is the wavelength.

\subsection{Model parameters of the Gabor filter and ResNet blocks}

For each wavelength, we use 64 Gabor filters with different orientations and scale values. The orientations take values with an additive interval of $\frac{\pi}{8}$ from $[0.75, 0.75 + \frac{\pi}{8}, 0.75 + \frac{\pi}{4}, 0.75 + \frac{3\pi}{8}, 0.75 + \frac{\pi}{2}, 0.75 + \frac{5\pi}{8}, 0.75 + \frac{3\pi}{4}, 0.75 + \frac{7\pi}{8}]$. The scales take values with a multiplicative interval of $\sqrt{2}$ from $0.75$ to $0.75 \times \sqrt{2}^7$.

In the ResNet structure shown in Figure \ref{model12}, each block comprises two sets of convolutional and batch normalization layers linked by ReLU activation \cite{maas2013rectifier}. The input is reintroduced to the output after the second batch normalization layer to guarantee reliable and resilient gradient computation. To begin the ResNet architecture, a single convolutional and batch normalization layer is employed to project the grayscale input image onto 64 channels. The subsequent three blocks continue to extract image features on 64 channels. In the subsequent four blocks, the results are projected to 128 channels. Following this, six blocks are used to capture higher-level features on 256 channels. The last three blocks extract global features on 512 channels. Following the completion of the ResNet blocks, the 512-channel image is projected back to grayscale using two convolutional layers. The purpose of increasing the number of dimensions is to allow the network to capture more complex and abstract features from the input image. With a higher number of channels, the convolutional layers can learn to detect more diverse patterns and structures within the image, leading to better performance.

\subsection{Data augmentation}

Data augmentation is a technique used to increase the size and diversity of a training dataset. In our model, we have applied a set of random transformations to the original data, including cropping, rotating, or adding noise. In active nematics experiments, free or extra microtubule filaments may float in the container, making the local orientation calculation difficult or inaccurate. To alleviate this, we proactively add random white lines in the input image to simulate the extra filaments. The transformed data is then added to the original dataset, increasing its size and variety, which helps to improve the performance of the machine learning model. Data augmentation can also help to reduce overfitting and make the model more robust to unseen data. By augmenting the training data with a variety of random transformations, we can help the model learn more generalizable features, which can improve its performance on new data.

\subsection{Tracking visualization over longer sequences}

With the matching algorithm discussed in Section 3.4, we can perform defect tracking over a long sequence of images.  Please refer to the supplementary material for a video of tracking results. 

\subsection{More supporting figures}
\begin{figure*}[!h]
    \centering
    \includegraphics[width=0.8\textwidth]{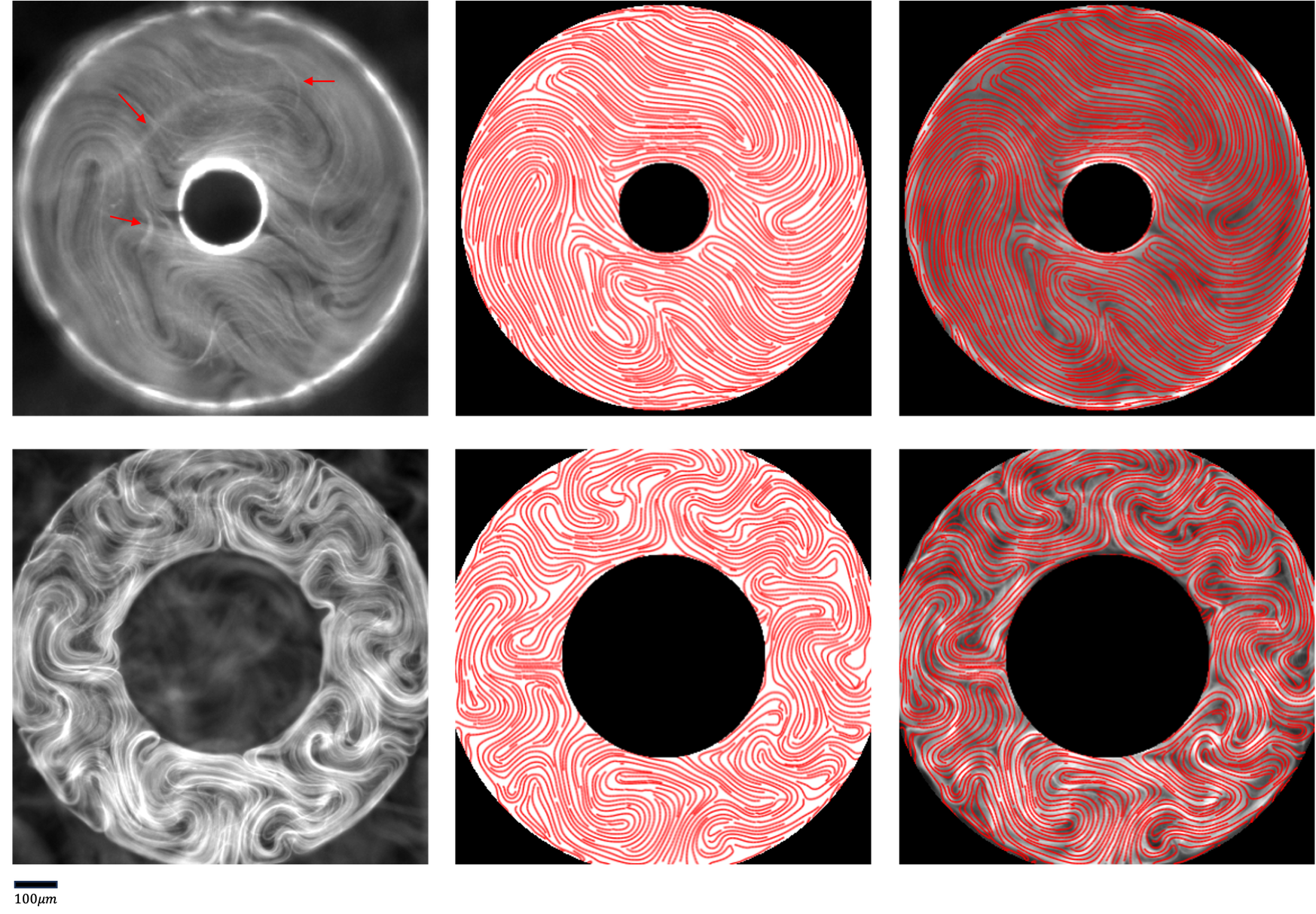}
    \caption{Machine Learning model performance on 2-dimensional fluorescence images with boundary conditions. The left, middle, and right columns are the input raw images, director fields extracted by our ML model, and director fields superimposed on raw images, respectively. Take the top row as an example: the red arrows on the raw image point to unbound microtubules drifting in the z-axis of the container. These microtubules do not correspond with the local orientation of the 2D nematic layer under investigation. Despite these extraneous elements, the machine learning model effectively computes the director field.}
    \label{annulus}
\end{figure*}

Figure \ref{annulus} illustrates the machine learning model's capability in processing fluorescence images of a 2D active nematic system with varying boundary conditions. The authors highlight that even in the presence of unbound microtubules floating along the z-axis of the 3D container, which is not aligned with the local orientation within the 2D nematic layer, the machine learning model adeptly filters out these elements, ensuring an accurate calculation of the director field.

\begin{figure*}[!h]
    \centering
    \includegraphics[width=0.8\textwidth]{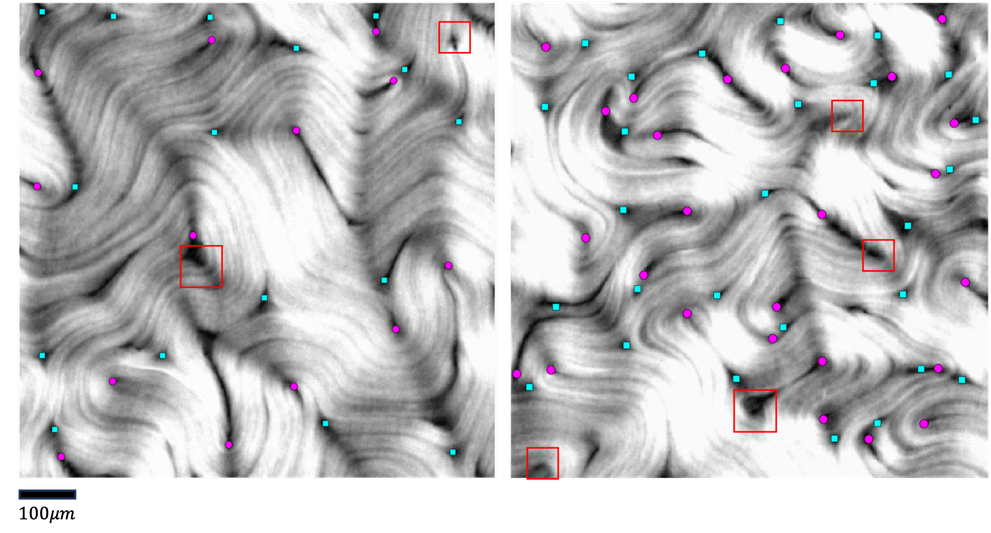}
    \caption{Some examples of missed defects using PolScope calculated director fields. The missed defects are highlighted by red boxes. This issue is due to noises in retardance images, such as under-/over- exposure, and exposure change; or it can occur in crowded regions where the defect density is high.}
    \label{polscope_defect}
\end{figure*}

Figure \ref{polscope_defect} showcases some examples in which some defects cannot be detected from the PolScope-calculated director field. We note that the local orientation by PolScope calculation is mostly accurate and the defect detection accuracy is generally good. The main contribution of this work is to develop a robust model that is less sensitive to noises in raw images and can be applied to experimental images from a variety of instruments.

\begin{figure*}[!h]
    \centering
    \includegraphics[width=0.6\textwidth]{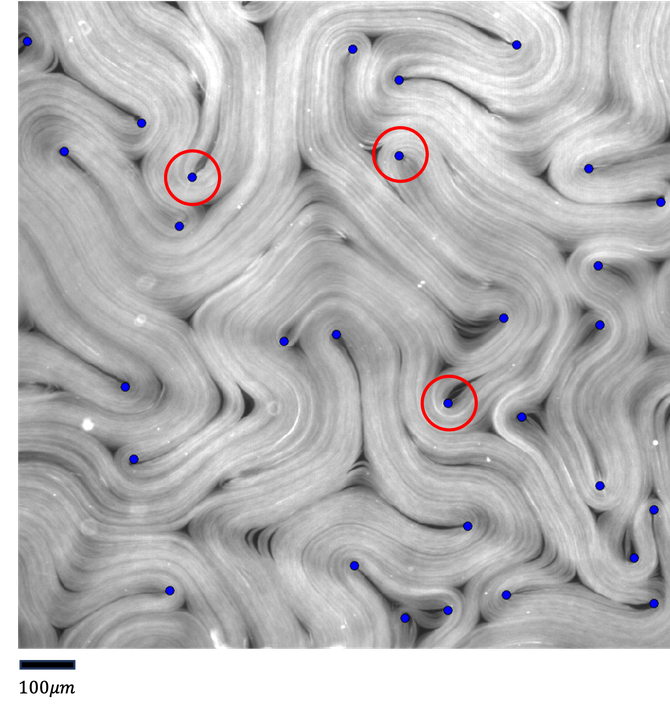}
    \caption{ Visualization of the 40-pixel threshold (red circle)  used in the evaluation of defect detection accuracy discussed in Section 3.3 and Figure 14.}
    \label{defectbox}
\end{figure*}

Figure \ref{defectbox} visualizes the 40-pixel threshold used to evaluate defect detection accuracy. 

\begin{figure*}[!h]
    \centering
    \includegraphics[width=0.9\textwidth]{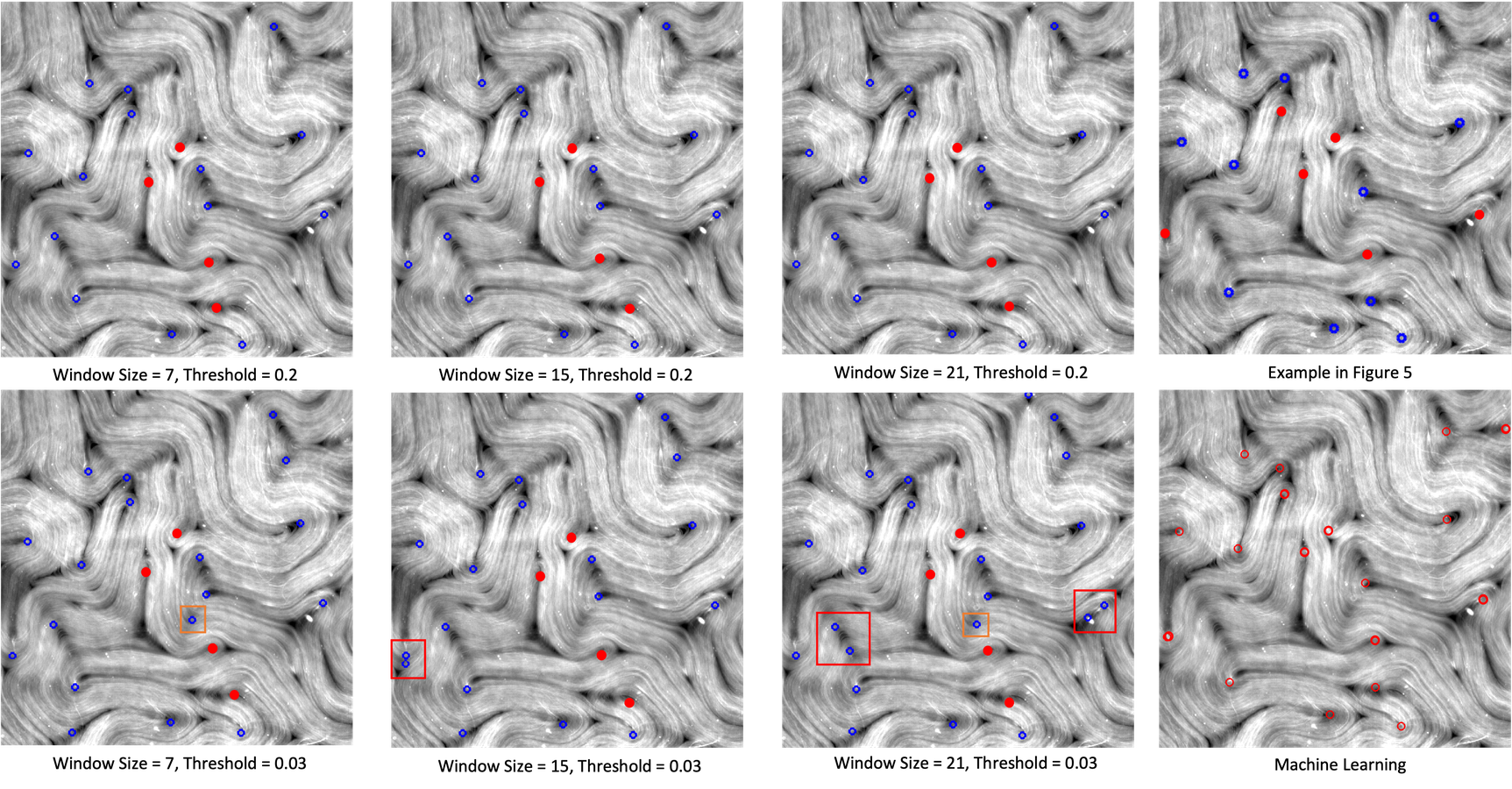}
    \caption{An extension to Fig. \ref{defect} to show how the results of the TM depend on the key parameter values of window size and defect detection threshold. We find that changes in the window size parameter may alter the defect detection results, but 4-5 defects remain undetected (solid red dots for each parameter setting) even for optimal values. Setting the defect detection threshold lower can lead to finding more defects, but also leads to multiple defects in one location (red box), or detection of non-existing defects (orange box).}
    \label{fig:changeParameters}
\end{figure*}

Figure \ref{fig:changeParameters} demonstrates the defect detection results using the TM generated director field. It is an extended analysis to Fig. \ref{defect}.

\section*{Conflicts of Interest}
There are no conflicts of interest to declare.

\section*{Acknowledgements}
This work was primarily supported by the Department of Energy (DOE) DE-SC0022291 (YL, PNT, YW, SF, MFH, PH). The experimental data acquisition and curation was supported by the Brandeis NSF MRSEC, Bioinspired Soft Materials, DMR-2011846 (ZZ). We also acknowledge computational support from the Brandeis HPCC which is partially supported by DMR-MRSEC2011486 and NSF OAC-1920147.

\clearpage

\newpage
\bibliography{main} 
\bibliographystyle{icml2021} 
\end{document}